\pdfoutput=1

\documentclass[11pt]{article}

\usepackage[]{acl}

\definecolor{bluex}{rgb}{0.27, 0.42, 0.81}
\definecolor{purplex}{HTML}{9564bf}
\definecolor{red3}{HTML}{C52A20}
\definecolor{red2}{HTML}{B36A6F}
\definecolor{red1}{HTML}{FFb5b5}
\definecolor{purple}{HTML}{B36A6F}
\definecolor{darkyellow}{HTML}{D5BA82}
\definecolor{blue1}{HTML}{508AB2}
\definecolor{blue2}{HTML}{C4E4E3}
\definecolor{green1}{HTML}{A1D0C7}
\definecolor{green2}{HTML}{BFF6BA}
\definecolor{green3}{HTML}{028100}
\definecolor{teal}{HTML}{508AB2}
\definecolor{purple1}{HTML}{8d3a94}

\usepackage{pifont}
\newcommand{\cmark}{\text{\ding{51}}}
\newcommand{\xmark}{\text{\ding{55}}}

\usepackage[listings]{tcolorbox}
\tcbuselibrary{listings,theorems}
\newtcolorbox{mybox}{colback=white!5!white,colframe=black!75!black, left=.05in, right=.05in}

\newtcbtheorem[number within=section]{exmp}{Example}%
{colback=green2!5,colframe=blue1,fonttitle=\bfseries, left=.02in, right=.02in,bottom=.02in, top=.02in}{exmp}
\newtcbtheorem[]{prompt}{Backward Prompting}%
{colback=green!5,colframe=green!35!black,fonttitle=\bfseries}{th}
\newtcbtheorem[number within=section]{thm}{Theorem}%
{colback=green!5,colframe=green!35!black,fonttitle=\bfseries}{th}
\newtcbtheorem[number within=section]{corr}{Corollary}%
{colback=green!5,colframe=green!35!black,fonttitle=\bfseries}{th}
\newtcbtheorem{method}{Method}%
{colback=green!5,colframe=green!35!black,fonttitle=\bfseries}{th}

\usepackage{times}
\usepackage{latexsym}

\usepackage[T1]{fontenc}

\usepackage[utf8]{inputenc}

\usepackage{microtype}

\usepackage{inconsolata}

\usepackage{graphicx}

\usepackage{caption} 

\usepackage{graphicx}
\usepackage{tabularx}
\usepackage{caption}
\usepackage{amsmath}%
\usepackage{MnSymbol}%
\usepackage{wasysym}%
\usepackage{lipsum,subcaption}
\usepackage{newfloat}
\usepackage{listings}
\usepackage{dsfont}
\usepackage{multirow} %
\usepackage{makecell}
\usepackage{amsfonts}
\usepackage{algorithm}
\usepackage{algpseudocode}
\usepackage{tabularx}
\usepackage {subcaption}

\usepackage{tikz}
\usetikzlibrary{shapes.misc,shadows}

\title{Towards a Unified Paradigm: Integrating Recommendation Systems as a New Language in Large Models}

\author{Kai Zheng\quad Qingfeng Sun  \quad  Can Xu\thanks{\quad  Corresponding author.}  \quad Peng Yu \quad Qingwei Guo\footnotemark[1] 
 \\
     Microsoft \\
      \texttt{\{zhengkai,qins,adrianyu,qingwei.guo\}@microsoft.com} 
            \texttt{nlpxucan@gmail.com}
      }

\begin{document}
\maketitle
\begin{abstract}

This paper explores the use of Large Language Models (LLMs) for sequential recommendation, which predicts users' future interactions based on their past behavior. We introduce a new concept, "Integrating Recommendation Systems as a New Language in Large Models" (RSLLM), which combines the strengths of traditional recommenders and LLMs. RSLLM uses a unique prompting method that combines ID-based item embeddings from conventional recommendation models with textual item features. It treats users' sequential behaviors as a distinct language and aligns the ID embeddings with the LLM's input space using a projector. We also propose a two-stage LLM fine-tuning framework that refines a pretrained LLM using a combination of two contrastive losses and a language modeling loss. The LLM is first fine-tuned using text-only prompts, followed by target domain fine-tuning with unified prompts. This trains the model to incorporate behavioral knowledge from the traditional sequential recommender into the LLM. Our empirical results validate the effectiveness of our proposed framework. 
\end{abstract}

\section{Introduction}

The field of sequential recommendation \cite{DBLP:journals/corr/abs-1905-01997,DBLP:journals/corr/abs-2001-04830} has long been focused on predicting users' future interactions with items based on their historical engagement sequences \cite{Hidasi2015SessionbasedRW,DBLP:journals/corr/abs-1808-09781,DBLP:journals/corr/abs-1809-07426}. This task is crucial for enhancing user experience and satisfaction in various online platforms, such as e-commerce, streaming services, and social media. The ability to accurately predict what a user will interact with next can significantly improve the relevance of recommendations, thereby increasing user engagement and retention.

Recently, the advent of Large Language Models (LLMs) \cite{zheng2023judging,touvron2023llama} has opened new avenues for sequential recommendation by conceptualizing it as a form of language modeling. This innovative approach leverages the powerful capabilities of LLMs to understand and generate human-like text, thereby offering a novel perspective on recommendation systems \cite{Bao_2023,cui2022m6rec,Dai_2023,geng2023recommendation}. LLMs, such as GPT-3 \cite{DBLP:journals/corr/abs-2005-14165} and BERT, have demonstrated remarkable proficiency in capturing complex patterns and relationships within textual data, making them well-suited for the task of sequential recommendation.

Traditional methods in sequential recommendation have typically represented items within LLMs' input prompts as either ID indices \cite{geng2023recommendation,Hua_2023} or textual metadata \cite{Bao_2023,cui2022m6rec,hou2023learning,Song2023SelfSupervisedMS,li2023text}. While these approaches have shown promise, they often fall short in encapsulating comprehensive world knowledge or demonstrating a deep understanding of user behavior. ID-based representations can be limited in their ability to convey rich semantic information about items, while textual metadata may not fully capture the nuances of user interactions and preferences.

To address these limitations, we propose a paradigm-shifting framework that integrates recommendation systems as a new language within large models, termed "Integrating Recommendation Systems as a New Language in Large Models" (RSLLM). The novelty of RSLLM lies in its unified prompting method, which combines ID-based item embeddings, learned by conventional recommendation models, with textual item features. By treating the "sequential behaviors of users" as a distinct language beyond text, RSLLM introduces a projector to align the traditional recommender's ID embeddings with the LLM's input space. This innovative approach allows the model to seamlessly incorporate behavioral knowledge from traditional sequential recommenders into the LLM, thereby enhancing its ability to predict user interactions more accurately.


Furthermore, we propose a two-stage fine-tuning framework for LLMs that introduces the alignment of user and item representations through a two-tower contrastive training approach. This framework refines a pretrained LLM using a combination of two contrastive losses and a language modeling loss. The initial stage of fine-tuning employs text-only prompts, aligning with the LLM's inherent language modeling capabilities. Subsequently, the framework undergoes target domain fine-tuning with unified prompts, effectively integrating behavioral knowledge from traditional recommenders. This two-stage process ensures the LLM's ability to leverage both textual and behavioral information, resulting in a robust and accurate recommendation system.
At the ID level, we align the traditional recommender's ID embeddings with the LLM's input space using a projector, effectively integrating user behavioral patterns. At the token level, the LLM processes textual item features, utilizing its extensive world knowledge to enhance the recommendation process. At the user/item level, we employ a two-tower contrastive learning method to seamlessly incorporate behavioral knowledge from the traditional sequential recommender into the LLM, enabling effective understanding and prediction of user behaviors based on their historical engagement sequences.

Empirical results \cite{Harper2016TheMD,DBLP:journals/corr/abs-1808-09781,10.1145/2043932.2044016} substantiate the efficacy of our proposed framework, demonstrating significant improvements in prediction accuracy and user satisfaction. By integrating recommendation systems as a new language within large models, RSLLM represents a significant step towards a unified paradigm in sequential recommendation, offering a novel and effective approach to capturing user behavioral patterns and world knowledge. This paradigm shift has the potential to revolutionize the field of recommendation systems, paving the way for more intelligent and context-aware recommendations that better serve users' needs and preferences.

Contributions of this work are three-fold:
(1) To the best of our knowledge, it is the first work to investigate the multi-granularity (ID, token, user/item) alignment of pre-trained large language models on sequential recommendation tasks. We explore this task on a sparse dataset that aligns with real-world scenarios, where recommendation systems (RS) and LLMs have distinct input formats.
(2) We introduce a novel framework, RSLLM, which is capable of aligning RS and LLM at multiple granularity levels.
(3) Our RSLLM approach outperforms state-of-the-art LLM sequence recommendation methods on various popular benchmarks.

\begin{table}\scriptsize 
\renewcommand\arraystretch{1}
\renewcommand\tabcolsep{1.8pt}
\centering
\begin{tabular}{l|c|c|c}
\hline
 \multicolumn{1}{l|}{Methods} &
 \multicolumn{1}{c|}{ID Level} &
 \multicolumn{1}{c|}{Token Level} &
\multicolumn{1}{c}{Item/User Level}
\\\cline{2-4}


\hline
{GRU4Rec} & \cmark &\xmark &\cmark \\
{Caser} & \cmark &\xmark &\cmark \\
{SASRec} & \cmark &\xmark &\xmark \\
{Llama2} & \xmark &\cmark &\xmark \\
{GPT-4}  & \xmark &\cmark &\xmark \\
{MoRec}  & \xmark &\cmark &\xmark \\
{TALLRec}  & \xmark &\cmark &\xmark \\
{LLaRA}  & \cmark &\cmark &\xmark \\
\hline
\textbf{RSLLM} & \cmark &\cmark &\cmark \\
\hline
\end{tabular}
\caption{\label{ablation-table} Comparison of different methods.}

\label{res:4}
   \vspace{-3em}  

\end{table}






\begin{figure*}[bht]
\centering
\includegraphics[scale=2, width=0.9\textwidth, trim=5 0 5 0,clip]{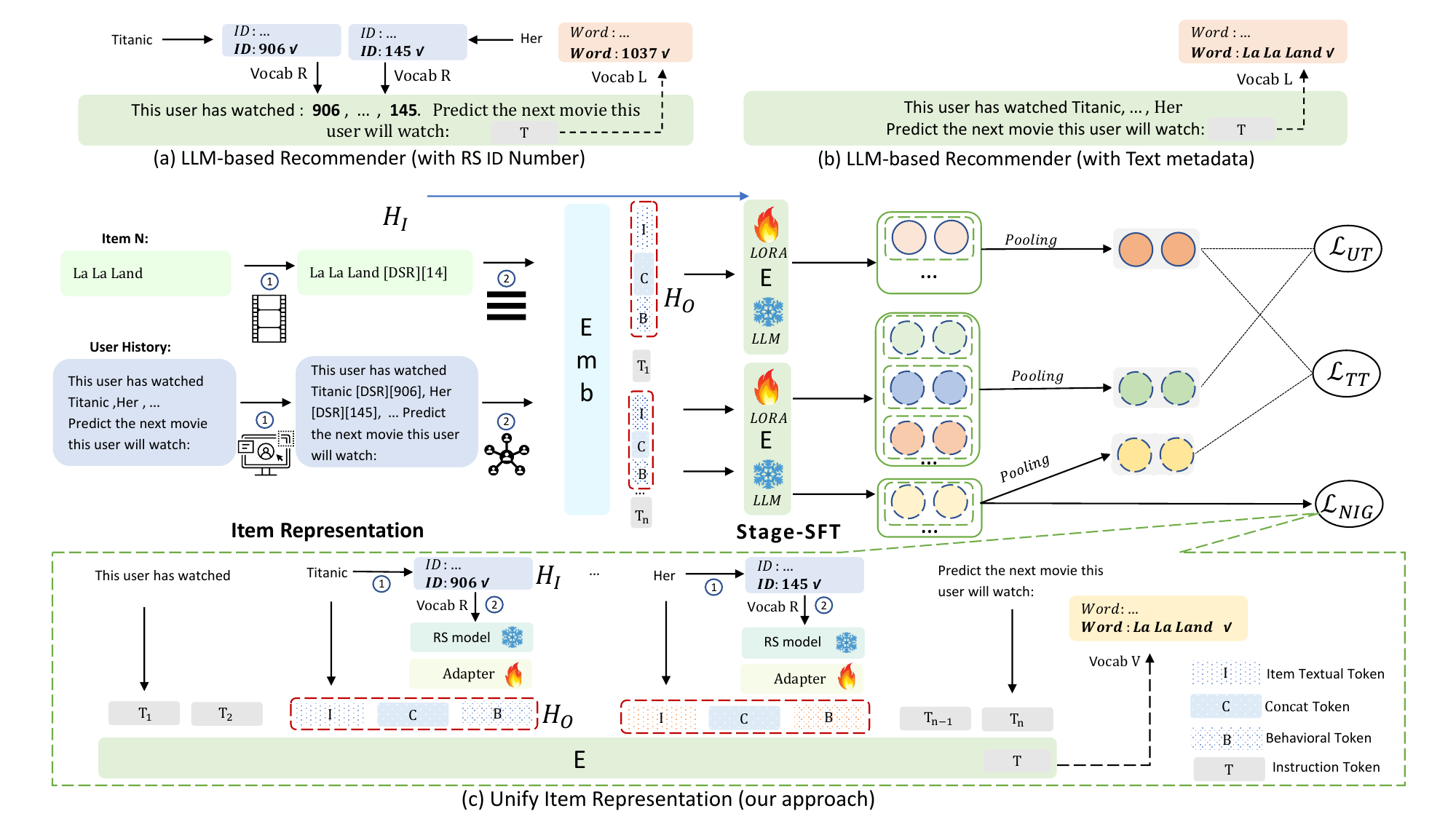}
\caption{An illustration of prior item representation methods and ours. (a) ID Number: represents an item with a numerical index. (b) Text Metadata: represents an item with its textual features, such as item title. (c) An illustration of our proposed RSLLM approach: integrates both textual tokens and behavioral tokens derived from the ID-based item embedding learned by traditional recommender models}
\label{fig:modelfig}
   \vspace{-1em}  
\end{figure*}

 
\section{Related Work}
\noindent \textbf{Large Language Models} LLMs like GPT-4 \cite{openai2024gpt4} and Llama \cite{touvron2023llama} have transformed fields such as natural language processing, machine learning, and information retrieval \cite{ouyang2022training,Zhao2023ASO, He2023AnnoLLMML, Xu2023WizardLMEL, Guo2023DrLI, Zheng2023AdversarialKS, Li2023EnablingPT}. Pretrained on extensive text and fine-tuned for specific tasks, LLMs excel at capturing complex patterns and relationships in textual data. They can generate human-like text and understand intricate semantic relationships
 \cite{Luo2023AugmentedLL,Ma2023FairnessguidedFP,Hu2023LLMAdaptersAA,Zhu2023MiniGPT4EV,wu2023bloomberggpt}. We utilize these capabilities to treat sequential recommendation as a form of language modeling.

\noindent \textbf{LLMs for Sequential Recommendation}
LLMs are a new direction in sequential recommendation systems \cite{hou2024large,li2023text,liu2023chatgpt,zhang2024generative,Zhao2024LetMD}. They address the limitations of traditional methods like collaborative filtering and content-based methods \cite{wu2022survey,Hidasi2015SessionbasedRW} by treating sequential recommendation as language modeling. 
Several studies represent items in prompts as either ID indices or textual metadata. For example, \citet{geng2023recommendation} use an ID number for each item, while \citet{Hua_2023} use a randomly-initialized ID token. Other models use textual metadata such as titles \cite{Bao_2023,cui2022m6rec} and descriptions \cite{hou2023learning,li2023text}. These models have improved recommendation accuracy.
However, existing approaches often fall short in encapsulating comprehensive world knowledge or demonstrating a deep understanding of user behavior. ID-based representations may not convey rich semantic information about items, and textual metadata may not fully capture the nuances of user interactions and preferences. To address these limitations, we propose RSLLM, a framework that introduces a unified prompting method combining ID-based item embeddings with textual item features. RSLLM treats "sequential behaviors of users" as a distinct language beyond text, incorporating behavioral knowledge from traditional sequential recommenders into the LLM. This enhances the LLM's ability to predict user interactions more accurately. We also propose a two-stage fine-tuning framework that aligns user and item representations through a two-tower contrastive training approach, leveraging both textual and behavioral information.

\section{Problem Formalization}

Task Formulation. Given a user $u \in \mathcal{U}$ who has an interaction sequence that consists of a sequence of $n$ items $U = {\{I_1, I_2, \dots , I_n\}}$ in chronological order, 
predict the next item $I_{n+1}$, where $n$ is the length of
$U$ and $I\in \mathcal{I}$. Each item $I$ has its corresponding ID and text
information (e.g. title, etc.).

\section{Methodology}
This section introduce the two important components in proposed RSLLM, including i) Unified Prompting Method; ii) Two-Stage Fine-Tuning Framework. Figure \ref{fig:modelfig} shows the overall RSLLM.

\subsection{Unified Prompting Method}

\subsubsection{ Prompt Construction}

The proposed method utilizes titles to describe items and uses item titles from historical interactions to describe users. Uniquely, to integrate collaborative information, we introduce additional user and item ID-related fields. 

\noindent \textbf{Text-ID Prompting.} 
For the ID numeric representation of the RS, we introduce the Text-ID Prompting approach for Large Language Model (LLM) instruction tuning. Items are represented via their textual metadata and ID numeric data within the prompt, as illustrated as follows (\textbf{Example 4.1}): 

\begin{exmp}{Text-ID prompt of RSLLM}{addcons-prompt}\small
		\textbf{\#Input Prompt\#:}
  This user has watched {\color{red3}Titanic [DSR] $[\textup{IID906}]$, City of Angels
[DSR] $[\textup{IID35}]$, .... Her [DSR] $[\textup{IID145}]$} in the previous. Please
predict the next movie this user will watch. The movie title
candidates are {\color{cyan} Avatar [DSR] $[\textup{IID3}]$, Schindler's List [DSR] $[\textup{IID78}]$,…,
Beastly [DSR] $[\textup{IID903}]$,… The Godfather [DSR] $[\textup{IID566}]$}, recommend one
movie for this user to watch next. The movie title you
recommend is\\

  \textbf{\#Output\#:}   {\color{teal}La La Land}
\end{exmp}

\noindent In this template, ``red font'' represents the list of item titles that the user interacted with, as a textual description of the user's preferences. ``Blue font'' refers to the title of the target item to be predicted, where the \texttt{<DSR>} token indicates that the following is a textual description of the recommendation ID, denoted by \texttt{<IID>}. The \texttt{<IID>} represents the IDs of the recommended items, to inject collaborative information. To maintain semantic consistency when integrating item IDs, we treat them as a feature of the item in the prompt.


\noindent \textbf{Hybrid Prompting.}
For the vector representation of the recommendation systems, we propose the Hybrid Prompting approach. This method integrates both textual and collaborative information from the recommendation systems into the prompts. It maintains the same prompt structure as Text-ID Prompting, but replaces the \texttt{<IID>} tokens with behavioral token representations obtained through the Hybrid Encoding module (\textbf{Example 4.2}).


\subsubsection{ Hybrid Encoding}
The Hybrid Encoding component is used to convert the input prompt into latent vectors, i.e., embeddings suitable for LLM processing. We employ a hybrid encoding approach, where for all textual content, we use the LLM's built-in tokenization and embedding mechanism to convert it into tokens and subsequent token embeddings. In contrast, when dealing with the item ID fields, we leverage an \textbf{Adapter} module, as illustrated in Figure \ref{fig:modelfig}(c), built with a conventional collaborative recommender, to extract collaborative information for the LLM to utilize.


Formally, for a prompt corresponding to the sample$(u,i)\in \mathcal{D}$, we first use the LLM Tokenizer to tokenize its textual content. The tokenization result is denoted as
$P = {\{t_1, t_2, \dots , t_n, i, \dots, t_K\}}$  where $t_k$ represents a text token, and $i$ signifies the item (ID) placed within the respective fields. We then further encode the prompt into a sequence of embeddings:
$E = {\{e_{t_1}, \dots , e_{t_k}, e_i, \dots, e_{t_K}\}}$,
where $e_{t_k}\in \mathcal{R} ^{1\times d}$ denotes the token embedding for $t_k$ in the LLM with dimension $d$, obtained via embedding lookup, and the embeddings for the item IDs, denoted as $e_i\in\mathcal{R} ^{1\times d}$, are obtained via the Adapter module.
 
To facilitate alignment, we project the ID-based item representation $e_s^i$ into the LLM space using a trainable projector, \textbf{Proj} (i.e., two-layer perceptions). This results in a behavioral token representation,  $\langle e_p^i \rangle = \text{Proj}(e_s^i; \theta_p)$, where $\theta_p$ are the parameters of the projector.

\noindent \textbf{Hybrid Token Representation.}
Upon obtaining the textual tokens $\langle e_t^i \rangle$ and the behavioral token $\langle e_p^i \rangle$ for item $i$, we integrate these two components. This integration provides a comprehensive representation of item $i$, effectively combining the distinct yet complementary aspects captured by each token. A specific concat $\langle e_t^c \rangle$ is used in this process (where the $c$ token indicates that the following subsequence is a representation of the recommendation ID):

\begin{align}
\langle e_h^i \rangle = \text{Concat}(\langle e_t^i \rangle; \langle e_t^c \rangle; \langle e_p^i \rangle)
\end{align}

\begin{exmp}{Hybrid prompt of RSLLM}{hybrid prompt}\small
		\textbf{\#Input Prompt\#:}
  This user has watched {\color{red3} Titanic [DSR] $[e_p^\textup{906}]$, City of Angels
[DSR] $[e_p^\textup{35}]$, .... Her [DSR] $[e_p^\textup{145}]$} in the previous. Please
predict the next movie this user will watch. The movie title
candidates are {\color{cyan} Avatar [DSR] $[e_p^\textup{3}]$, Schindler's List [DSR] $[e_p^\textup{78}]$,…,
Beastly [DSR] $[e_p^\textup{903}]$,… The Godfather [DSR] $[e_p^\textup{566}]$} , recommend one
movie for this user to watch next. The movie title you
recommend is\\

  \textbf{\#Output\#:}   {\color{teal}La La Land}
  
\end{exmp}

\begin{algorithm}
\caption{Two-Stage Optimization Algorithm}\label{alg:cap}
\begin{algorithmic}[1]
\Require $s$: number of training iterations
\State $\mathcal{D}: \textup{dataset}$
\State $M: \textup{model}$
\State $N \gets 1$
\State ${H}_{I} \gets \textup{GEN}(\mathcal{D}, I)$ \Comment Text-ID Prompting 
\State ${H}_{O} \gets \textup{GEN}({H}_{I}, O)$ \Comment Hybrid Prompting   

\Comment Stage 1: Text-Only Fine-Tuning
\For{$i = 1$ to $s$}
    \State Update $M$ by minimizing the loss function in Eq.\ref{eq:7} on ${H}_{I}$
\EndFor

\Comment Stage 2: Target Domain Fine-Tuning
\For{$i = 1$ to $s$}
    \State Update $M$ by minimizing the loss function in Eq.\ref{eq:7} on ${H}_{O}$
\EndFor
\State \textbf{return} $M$

\end{algorithmic}
\end{algorithm}

\subsection{Two-Stage Fine-Tuning Framework}

Our Two-Stage Fine-Tuning Framework refines a pretrained LLM using two contrastive losses and a language modeling loss. It consists of two stages:

\noindent \textbf{Text-Only Fine-Tuning} The LLM is fine-tuned using text-only prompts, aligning with its inherent language modeling capabilities. Items are represented via their textual metadata, allowing the LLM to better understand and interpret the textual features of the items.

\noindent \textbf{Target Domain Fine-Tuning} A subsequent round of fine-tuning is performed with unified prompts, integrating ID-based item embeddings from conventional recommendation models with textual item features. This trains the model to incorporate behavioral knowledge from traditional sequential recommenders.

By sequentially fine-tuning the LLM, the model can understand and interpret both the textual features of items and the behavioral patterns of users, resulting in a more robust and effective RS.  To strike a balance between efficiency
and efficacy, we conduct LoRA \cite{hu2021lora} tuning as introduced in for the LLM, while training the projector at the same time. The training objective remains the same in both stages.

\subsubsection{ Contrastive Alignment}
We now delve into the methodology for optimizing the model's parameters. The primary objective we employ for fine-tuning Large Language Models (LLMs) is the Next Item Prediction (NIP) Objective. The NIP objective is designed to predict the textual description of the subsequent item based on the historical sequence of items described in text. Let us represent the tokenized user sequence as 
as$(u_1, u_2, \dots , u_{n})\}$, where $n$ denotes the length of the sequence. The first 
$m$ tokens correspond to all items except the last one, with the remaining 
$n-m$ tokens dedicated to the target item. Our proposed objective for adapting LLMs to sequential recommendation is formulated as follows:
\begin{align}
\mathcal{L}_\mathcal{N} & =  -\mathbb{E}\sum_{j=m+1}^{n}[\textup{log}_{\mathcal{M}_\theta} (P(u_j|u_{1:j-1};\theta))] \
\end{align}
where $\theta$ encompasses all trainable parameters within the LLM.

To address the limitation of NIP, which operates on a token level rather than an item/user level, we introduce Contrastive Alignment. This involves an auxiliary contrastive objective that functions at the item/user level. We employ a two-tower training framework: one tower processes the target item, yielding a mean-pooled feature 
${\textup{g}^\textup{I}}$, while the other incorporates the entire user sequence, resulting in features 
${\textup{g}^\textup{U}}$ for the user history and 
${\textup{g}^\textup{I|U}}$ for the target item conditioned on the user history. We experiment with two contrastive losses for user- and item-level alignments, both inspired by the InfoNCE loss, a robust choice in contrastive learning \cite{DBLP:journals/corr/abs-1807-03748, li-etal-2022-improving, li-etal-2023-translation,gao-etal-2021-simcse,pmlr-v119-chen20j}:
\begin{align}\label{eq:3}
\mathcal{L}_\mathcal{I} & =  -\frac{1}{N}\sum_{i=1}^{N}\textup{log} \frac{{\textup{e}^{(\textup{cos}(\textup{g}_i^{I|U}, \textup{g}_i^I)/{\tau_c})}}}{\sum_{j=1}^{N}{\textup{e}^{(\textup{cos}(\textup{g}_j^{I|U}, \textup{g}_i^I)/{\tau_c})}}} \
\end{align}

\begin{align}\label{eq:4}
\mathcal{L}_\mathcal{U} & =  -\frac{1}{N}\sum_{i=1}^{N}\textup{log} \frac{{\textup{e}^{(\textup{cos}(\textup{g}_i^{U}, \textup{g}_i^I)/{\tau_c})}}}{\sum_{j=1}^{N}{\textup{e}^{(\textup{cos}(\textup{g}_j^{U}, \textup{g}_i^I)/{\tau_c})}}} \
\end{align}
where in-batch negative samples are used, 
${N}$ represents the batch size, 
cos(·, ·) denotes the cosine similarity, and 
${\tau_c}$ is the temperature parameter for contrastive alignments. Our final training objective combines 
$\mathcal{L}_\mathcal{N}$ with the contrastive losses as depicted in 
 Figure\ref{fig:modelfig}c:

 \begin{align} \label{eq:7}
\mathcal{L} =  \mathcal{L}_\mathcal{N} + \gamma \mathcal{L}_\mathcal{I} + \beta \mathcal{L}_\mathcal{U}
\end{align}

\begin{table*}\scriptsize  
\renewcommand\arraystretch{0.81}
\renewcommand\tabcolsep{1.6pt}
\centering
\begin{tabular}{l |c c  |c  c   |c  c |c c}
\hline

\multirow{2}{*}{Models} &
\multicolumn{2}{c|}{MovieLens 
} &
\multicolumn{2}{c|}{Steam}  
 &
\multicolumn{2}{c|}{LastFM} & 
\multicolumn{2}{c}{Ave.}  \\\cline{2-9}


 & ValidRatio&{HitRatio@1}      & ValidRatio &{HitRatio@1}   & ValidRatio &{HitRatio@1} & ValidRatio &{HitRatio@1} \\

\hline
\multicolumn{1}{c}{\textup{$Traditional$}}\\
 \hline

GRU4Rec   &1.0000 &0.3750 
 &1.0000 &0.4168 
  &1.0000 &0.2616 &1.0000 &0.3511
 \\
   Caser &1.0000 &0.3861 
 &1.0000 &0.4368  
  &1.0000 &0.2233 &1.0000 &0.3487\\

     SASRec&1.0000 &0.3444
  &1.0000 
  &0.4010      &1.0000 
  &0.2233 &1.0000  &0.3229  \\  
\hline

\multicolumn{1}{c}{\textup{$LLM-based$}}\\
 \hline
Llama2 &0.4421 &0.0421 
 &0.1653 &0.0135
 &0.3443 &0.0246 &0.3172 &0.0267
 \\
    GPT-4 &0.9895 &0.2000 
 &0.9798 &0.3626 
  &1.0000 &0.3770  &0.9897 &0.3132\\

     MoRec&1.0000 &0.2822 
  &1.0000 
  &0.3911 
    &1.0000 
  &0.1652 &1.0000  &0.2795\\

  TALLRec&0.9263 &0.3895  
 &0.9840  &0.4637 
 &0.9836  &0.4180 &0.9646 &0.4237
 \\
  LLaRA (GRU4Rec)& {0.9684} & {0.4421} 
& {0.9975} & {0.4924} 
& {0.9836} & {0.4344} & {0.9831} &0.4563
 \\
  LLaRA (Caser)& {0.9684} & {0.4737} 
& {0.9966} & {0.4874} 
& {0.9918} & {0.4344} & {0.9856} &0.4651
 \\
   LLaRA (SASRec)& {0.9684} & {0.4421}
& {0.9975} & {0.4949} 
& {1.0000} & {0.4508} & {0.9886} &0.4626 
 \\
\hline

\multicolumn{1}{c}{\textup{$Ours$}}\\
 \hline
 
RSLLM (GRU4Rec)&0.9698 &0.4947  &0.9980
 &\underline{0.5130} 
&0.9919
 &\underline{0.4649} &0.9865 &0.4908
 \\
   RSLLM (Caser)  &0.9701 &\textbf{0.5273}  
 &0.9968 &0.4953  
 &0.9939 &{0.4646} &0.9869 &\underline{0.4957} \\
   RSLLM (SASRec)&0.9700 &\underline{0.5005} 
  &0.9982 
  &\textbf{0.5241}   
  & {1.0000} & \textbf{0.4980} & {0.9894} &\textbf{0.5075}\\ 

\hline

\end{tabular}
\caption{\label{font-table}  The Results of RSLLM compared with traditional sequential recommender models and LLMs-based methods. \textbf{Bold} and \underline{underline} are the significant best and the second-best results compared to the \textbf{Baseline} model (paired student’s t-test with $p$-value $<$ 0.05, Ave. stands for average result).
}
\label{result:table_1}
    \vspace{-1em}  

\end{table*}

\section{Experiments}
This section first introduces the experimental settings in Section \ref{exp:setup}, and then presents the main experimental results in Section \ref{exp:5.2}. Ablation studies were conducted in Section \ref{exp:5.3}. In Section \ref{exp:5.4}, we compare RSLLM with different representation and recommendation models.

\subsection{Experimental Setup}\label{exp:setup}
Following the experiment setting in LLaRA \cite{liao2024llara}, we conduct experiments on three real-world datasets:MovieLens \cite{Harper2016TheMD}, Steam \cite{DBLP:journals/corr/abs-1808-09781}, and LastFM \cite{10.1145/2043932.2044016}, detailed statistics of the datasets are provided in Table \ref{stats-table}. For each benchmark, we conduct experiment following \cite{liao2024llara}: For each sequence, we randomly select 20 non-interacted items to construct the candidate set, ensuring the inclusion of the correct subsequent item. RSLLM and other baseline models aim to identify the correct item from this candidate set, and their performance is evaluated using the HitRatio@1 metric.  And LLM-based  metric: valid ratio. It quantifies the proportion of valid responses (i.e., items in the candidate set) across all sequences, serving as a measure of the models’ capability of instruction following.
We repeated the experiment 5 times and averaged the results according to the previous works \cite{DBLP:journals/corr/abs-1808-09781,Harper2016TheMD}. The \textbf{Baseline} models include both traditional sequential recommender models such as GRU4Rec \cite{DBLP:journals/corr/HidasiKBT15}, Caser \cite{DBLP:journals/corr/abs-1809-07426}, and SASRec \cite{DBLP:journals/corr/abs-1808-09781}, and LLM-based Models such as Llama2 \cite{touvron2023llama}, GPT-4 \cite{openai2024gpt4}, MoRec \cite{yuan2023recommender}, ,TALLRec \cite{Bao_2023}  and MetaST (Wang et al., 2021a),
a SOTA LLM-based sequence recommendation method.


\begin{table}\scriptsize 
\renewcommand\arraystretch{1}
\renewcommand\tabcolsep{1.8pt}
\centering
\begin{tabular}{l|c|c|c}
\hline
 \multicolumn{1}{l|}{Dataset} &
 \multicolumn{1}{r|}{MovieLens} &
 \multicolumn{1}{c|}{Steam} &
\multicolumn{1}{r}{LastFM}
\\\cline{2-4}


\hline
{\# Sequence} & 943 &11,938 &1,220 \\
{\# Item} &  1,682 & 3,581 &4,606 \\
{\# Interaction} & 100,000 & 274,726 &73,510 \\
{\# Sparsity } & 93.7\% & 99.4\% &98.7\% \\
\hline
\end{tabular}
\caption{\label{stats-table} Statistics of all datasets.}

\label{res:4}
    \vspace{-2em}  
\end{table}

\begin{table*}\scriptsize  
\renewcommand\arraystretch{0.81}
\renewcommand\tabcolsep{1.6pt}
\centering
\begin{tabular}{l |c c|c  c |c  c}
\hline

\multirow{2}{*}{Models} &
\multicolumn{2}{c|}{MovieLens 
} &
\multicolumn{2}{c|}{Steam}  
 &
\multicolumn{2}{c}{LastFM} \\\cline{2-7}


 & ValidRatio&{HitRatio@1}    & ValidRatio &{HitRatio@1}  & ValidRatio &{HitRatio@1}   \\

\hline

Baseline$\clubsuit$   &0.9684 &0.4421 &0.9975  
 &0.4949 &1.0000 &0.4508 
 \\
   RSLLM &0.9700 &0.5005 &0.9982 &0.5241 &1.0000 &0.4980  \\

\hline

\multicolumn{1}{c}{\textup{$Ablation \ For \ Item\ Representation\ $}}\\
 \hline

w/o  Textual Feature (Titles,etc.) Representation$\spadesuit$ &0.9369 &0.4010 &0.9501
 &0.4201 &0.9135 &0.2230 

 \\

 w/o Item ID (IID/Embedding) Representation $\dag$&0.9421 &0.4152 &0.9735
  &0.4794 
  &0.9811 &0.4317

 \\ 

 w/o IID Tokens&0.9669 &0.4316 &0.9801
  &0.4890 
  &0.9811 &0.4347

 \\ 

  w/o Pre-loading item Embeddings.&0.9608 &0.4343 &0.9866
  &0.4905 
  &0.9818 &0.4388

 \\

  \hline
\multicolumn{1}{c}{\textup{$Ablation \ for\ User-Item\ Alignment $}}\\
 \hline

  RSLLM (U-I Only) &0.9686 &{0.4754} &0.9980 
  &0.5005 
  &{1.0000} &0.4849
  \\ 
RSLLM (I-I Only) &0.9695 &{0.4701} &0.9978 
  &0.4989
  &{1.0000} & 0.4831
  \\ 

RSLLM (w/o. Contrastive Alignment )&0.9675 &{0.4544} &0.9978 
  &0.4958 
  &{0.9970} & 0.4765
  \\ 
  \hline
\multicolumn{1}{c}{\textup{$Different\ Learning\ Strategies$}}\\
 \hline

  Stage1 only \dag &0.9608 &0.4073 &0.9866
  &0.4905 
  &0.9818 &0.4388

 \\ 
Stage2 only \ddag &0.9684 &0.4231 &0.9962  &0.5148
 &{1.0000} &{0.4974}

 \\

   RSLLM (Two-stage) &{0.9700} &{0.5005} &{0.9982} &{0.5241} &{1.0000} &{0.4980}  \\

\hline

\end{tabular}
\caption{\label{font-table} The ablation result over MovieLens, Steam and LastFM. w/o. denotes that
we only remove one component from RSLLM. 
$\clubsuit$ results taken from \cite{liao2024llara}.
}
\label{res:auto_3}
    \vspace{-1em}  

\end{table*}

\noindent \textbf{Implementation Details}
Our
method uses a task description prompt/template (primary prompt) for the prediction of each task. We use
Adam optimizer with learning rate 1e-5, warm-up
rate of 0.1 and weight decay 1e-3 in training. The embedding dimension $d$ is 64. We set 256 as batch size. We train RSLLM maximum of 3 epochs  in Stage 1, and then further fine-tune the model for Stage 2.
The $\gamma$, $\beta$ in Eq.\ref{eq:7} are 0.3, 0.4. and adopt ${\tau_c}$ = 0.5 (see Eq.\ref{eq:3} and Eq.\ref{eq:4}).  Early stopping on validation is adopted as a regularization strategy. All the hyper parameters are determined by grid search.
To mitigate
the impact of randomness, we report the average outcomes of five
runs using different random seeds.

 \begin{figure}[!htb]
\centering
     \includegraphics[width=0.5\textwidth, scale=2, trim=20 43 10 30,clip]{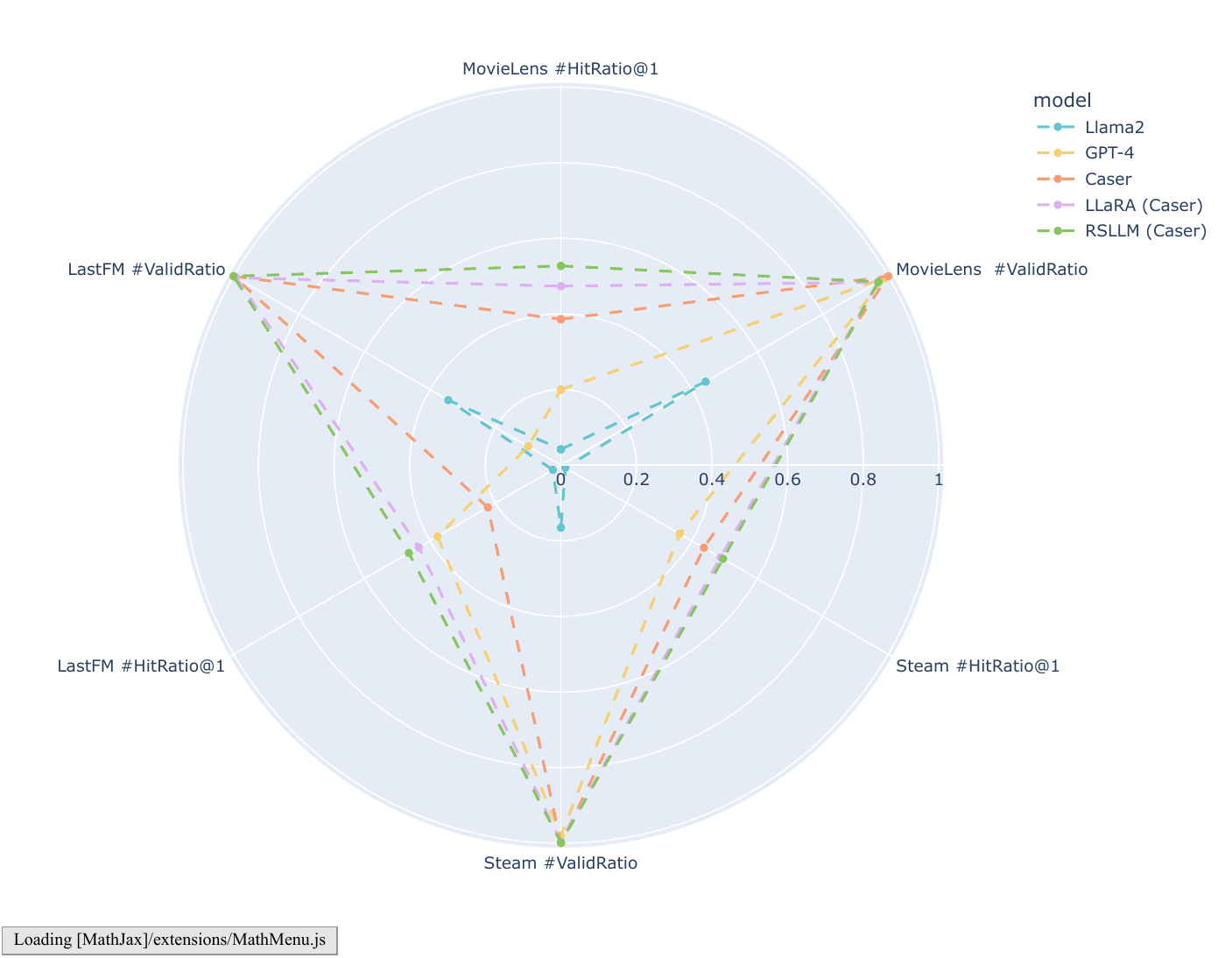}
     \caption{Results of recommendation model efficiency analysis. We compare RSLLM with strong baselines with Caser backbone. The GRU4Rec and SASRec result are presented in Appendix \ref{A:1}.}
     \label{fig:3}
         \vspace{-2em}  

\end{figure}

\subsection{Main Results}\label{exp:5.2}

As shown in Table \ref{result:table_1}, our proposed RSLLM significantly outperforms all baseline models across the three datasets in terms of HitRatio@1 and ValidRatio metrics. Specifically, on the MovieLens dataset, RSLLM surpasses the best baseline by 4.3\% in HitRatio@1. Similar trends are observed on the other two datasets, with RSLLM exceeding the next best method by 3.0\% on Steam and 4.8\% on LastFM in HitRatio@1.

The traditional sequential recommendation models, such as GRU4Rec, Caser, and SASRec, yield lower HitRatio@1 scores compared to RSLLM. These models generate predictions solely based on users' behavioral patterns, without incorporating any semantic information about the items. This highlights the importance of integrating world knowledge about items into the recommendation process.

\noindent \textbf{Performance of LLM-based Methods} Regarding LLM-based methods, two key observations emerge from the results. Firstly, the underperformance of the standalone LLMs (i.e., Llama2 and GPT-4) emphasizes the need to adapt these models to recommendation tasks to boost their performance in this domain. Secondly, the LLM4Rec methods (i.e., MoRec and TALLRec) show slight improvements over the standalone LLM methods, but their recommendation capability, as indicated by the HitRatio@1 metric, is still inferior to that of RSLLM. This highlights the necessity for a comprehensive approach that combines the strengths of both LLMs and traditional recommendation models.

\noindent \textbf{Validity of Recommendations}
RSLLM achieves a high validity ratio exceeding 96.9\% across all datasets, demonstrating the model's proficiency in adhering to instructions when generating recommendations. It is worth noting that all generative methods that incorporate LLMs might produce invalid answers. For instance, the backbone LLM of RSLLM, Llama2, only achieves valid ratios of 0.4421, 0.1653, and 0.3443 on the MovieLens, Steam, and LastFM datasets, respectively. The substantial improvement in valid ratios by RSLLM can be attributed to the fact that RSLLM has been instruction-tuned on the sequential recommendation task. This underscores the effectiveness of instruction-tuning in enhancing the validity of recommendations.

The results demonstrate the superiority of RSLLM in sequential recommendation tasks compared to both traditional and LLM-based baselines. RSLLM's strong performance can be attributed to its ability to effectively incorporate item-level semantic information and its instruction-tuning on the recommendation task, which enhances the validity of the generated recommendations. These findings highlight the potential of combining the strengths of LLMs and traditional recommendation models to achieve state-of-the-art performance in sequential recommendation.

\subsection{Ablation Study}\label{exp:5.3}
\noindent \textbf{Item Representation} As shown in Table \ref{res:auto_3}, without Textual Feature Representation, we directly fine-tune the model on behavioral data without using textual features such as titles and descriptions.
Without Item ID Representation, we remove the item ID representations, which are crucial for capturing the unique identity and sequential relationships of items.
Without IID Tokens, we limit the model by not using the <IID> tokens, which typically represent the IDs of the items in the dataset. Without Pre-loading Item Embeddings, we disregard the preloaded embeddings that are typically used to inject prior knowledge about items into the model. As shown in the results, the fully fine-tuned PLMs without textual feature representation perform worse than our proposed RSLLM method  (16.8\% HitRatio@1 lower average, especifically drop 27.5\% HitRatio@1 and 9.7\% ValidRatio in LastFM ), showing the positive contribution of textual features for accurate recommendations. Further, removing item ID representation or IID tokens also delegate the performance  by 7.4\% and 6.4\% average, showing the importance of using these components to learn a reasonable item representation. Similarly, without pre-loading item embeddings, the model achieves similar performance as when the embeddings are included. It is recommended to directly train the item representation parameters.

\noindent \textbf{Contrastive Alignment} Next, we show the effect of different user-item alignment strategies in our RSLLM framework. The RSLLM (U-I Only) and RSLLM (I-I Only) configurations only retain the user-tower to target-tower alignment and the target-tower to target-tower alignment, respectively.
As shown in Table \ref{res:auto_3}, these two aligments successfully boost up the model performance(2.3\% HitRatio@1 gain average compare with Basline, especifically in MovieLens and LastFM ). However, their corresponding models perform worse than the ones supported by the full RSLLM. This shows that aligment from different views provide meaningful and different training signals to the models. 
Interestingly, models trained on the Target-Tower to User-Tower Alignment (I-I Only) perform better than the ones trained on the User-Tower to Target-Tower Alignment (U-I Only), indicating that the reciprocal alignment from the user-tower to the target-tower might be more instrumental in capturing the bidirectional relationship between user history and item preferences.
Finally, the trained model removing Contrastive Alignmen perform worse than RSLLM (3.1\% HitRatio@1 lower average), showing the importance of Contrastive Alignment.

\noindent \textbf{Impact of Two-stage Training}
Finally, we examine the effect of the two-stage training framework in our RSLLM model. In Table \ref{res:auto_3}, we show the model performance with only the first stage (Stage 1 Only), only the second stage (Stage 2 Only), and with both stages (RSLLM Two-stage). The two-stage training has an important effect on the model performance. Without integrating behavioral knowledge, the performance gap almost disappears. The sequential fine-tuning also has a positive effect on the model performance. 
In particular, in the MovieLens dataset, the model performance increases significantly after the second stage. This shows that the initial text-only fine-tuning provides a necessary foundation for the subsequent target domain fine-tuning to be most effective. 
The results validate the hypothesis that a phased approach, which first establishes a strong textual understanding and then refines this with behavioral knowledge, leads to the most accurate and robust sequential recommendation model. 

\subsection{Discussions}\label{exp:5.4}
\begin{figure}[!htb]
\vspace{-1em}  
\centering
     \includegraphics[width=0.5\textwidth, scale=2, trim=20 23 20 0,clip]{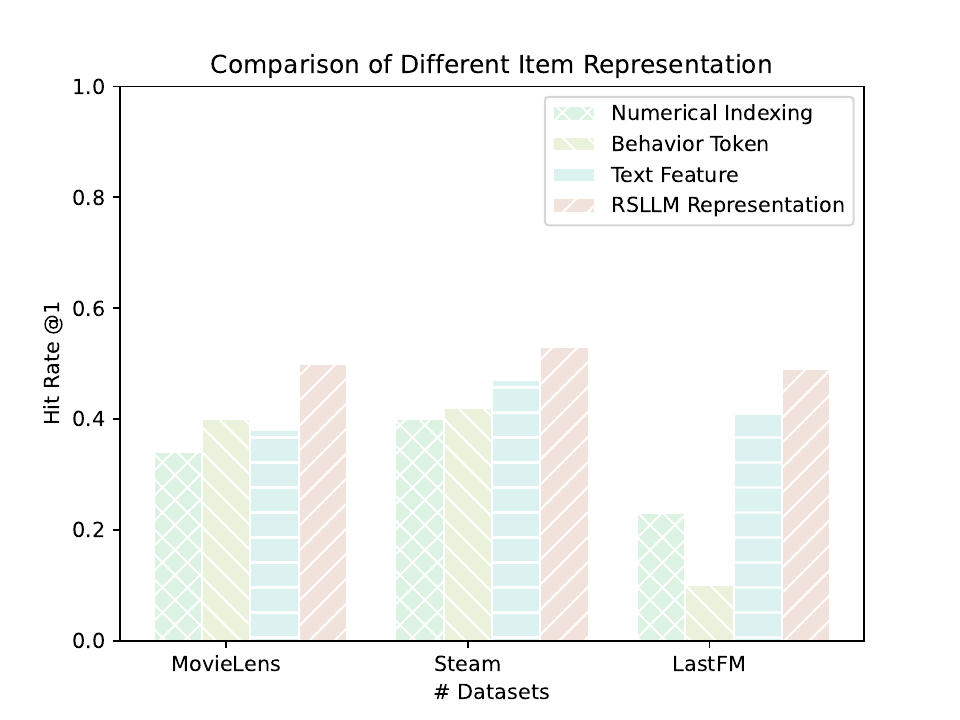}
     \caption{The performance comparison of different item representation methods (i.e., numerical index, behavioral token,
textual feature, LLaRA and RSLLM representation) in datasets: MovieLens, Steam and LastFM.}
     \label{fig:2}
              \vspace{-1em}  

\end{figure}

\noindent \textbf{Evaluation of item representation methods} 
We conduct a comprehensive empirical evaluation of prevalent item representation methods in sequential recommendation tasks, including numerical indexing, behavior tokens, text feature representation, and the unified representation utilized in our RSLLM framework.
As shown in Figure \ref{fig:2}, the results demonstrate that the item representation method employed by RSLLM outperforms the other approaches in terms of HitRatio@1 across all three datasets. This not only validates the effectiveness of RSLLM's innovative item representation technique, but also highlights the advantages of its more comprehensive alignment between the LLM and the recommendation system, compared to conventional single-faceted item representation methods.

Regarding the limitations of the individual methods:
Numerical Indexing: For numerical indexing, LLMs do not initially store any inherent information. These indices are processed as plain text by the LLMs, causing the tokenizer to divide them into multiple tokens, which may limit the model's understanding.
Behavior Token: When using behavior tokens, LLMs primarily exploit the distribution of the input behavioral embeddings, without effectively extracting the knowledge encapsulated within the LLMs.
Text Feature: In the case of text features, the absence of user behavior patterns allows the LLM to solely infer the correlations among items in a user's historical interactions, guided solely by the background knowledge of these items preserved in the LLM.
In contrast, RSLLM's unified item representation integrates both world knowledge and sequential information, thereby enhancing performance in sequential recommendation. By fusing item ID, behavior tokens, and text tokens, RSLLM is able to capture a more comprehensive representation of items, which leads to superior recommendation capabilities compared to the other approaches.

\noindent \textbf{Discussion for Different recommendation model}
We evaluate our proposed RSLLM framework using item embeddings derived from three traditional sequential recommendation backbones: GRU4Rec, Caser, and SASRec. These models represent the three main categories of recommendation models : RNN-based, CNN-based and self-attention-based. 

The empirical results in Figure \ref{fig:3} and Appendix \ref{A:1} show that RSLLM outperforms all baseline models(RS, GPT-4, LLM-based model) and using SASRec as the backbone achieves the best performance, outperforming the other backbones. This validates that self-attention is better able to capture both local and global dependencies in user-item interactions compared to RNNs and CNNs.
However, the performance gains of SASRec over GRU4Rec and Caser are marginal in some cases, indicating the sequential patterns captured by different backbones do not vary dramatically. 
The key factor for RSLLM's improved performance is the unified prompting and fine-tuning approach not backbone.

\section{Conclusion}

In this paper, we introduce a novel and effectiveness framework: RSLLM. 
Experiments on various benchmarks show the effectiveness of RSLLM. In the future, we plan to expand RSLLM to other recommendation tasks, like conversational recommendation and multi-modal recommendation.



\clearpage

\section*{Limitations}
The RSLLM fine-tuning process and the integration of ID-based item embeddings with textual item features can be computationally intensive, which requires large GPU memory.


\bibliography{custom}
\clearpage

\appendix




\section{Appendix}
\label{sec:appendix}
\subsection{GRU4Rec and SASRec result in Figure 3}\label{A:1}
\begin{figure}[!htb]
\centering
     \includegraphics[width=0.5\textwidth, scale=2, trim=20 43 10 30,clip]{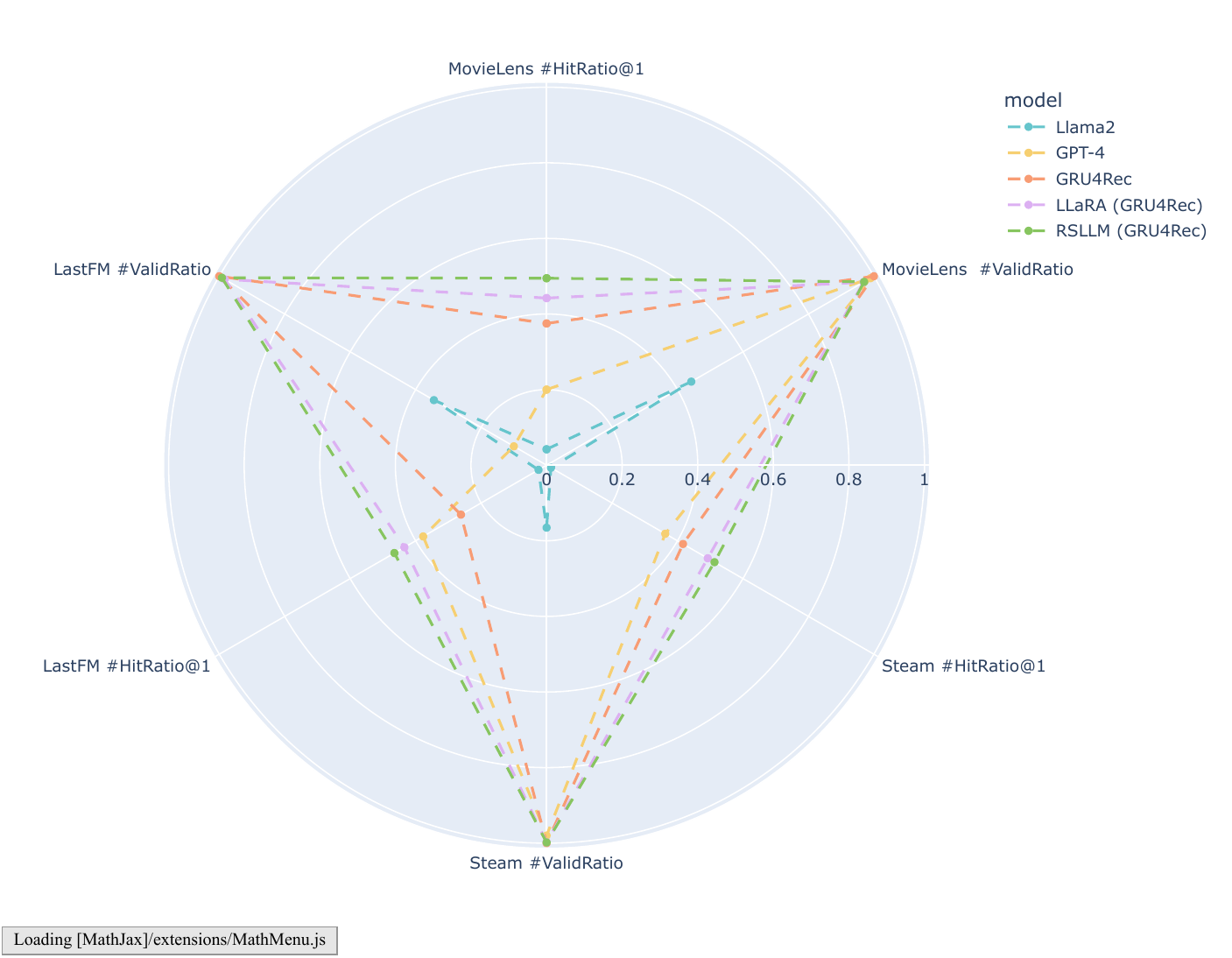}
     \caption{Results of recommendation model efficiency analysis. We compare RSLLM with strong baselines with GRU4Rec backbone.}
     \label{fig:4}
         \vspace{-2em}  

\end{figure}

\begin{figure}[!htb]
\centering
     \includegraphics[width=0.5\textwidth, scale=2, trim=20 43 10 30,clip]{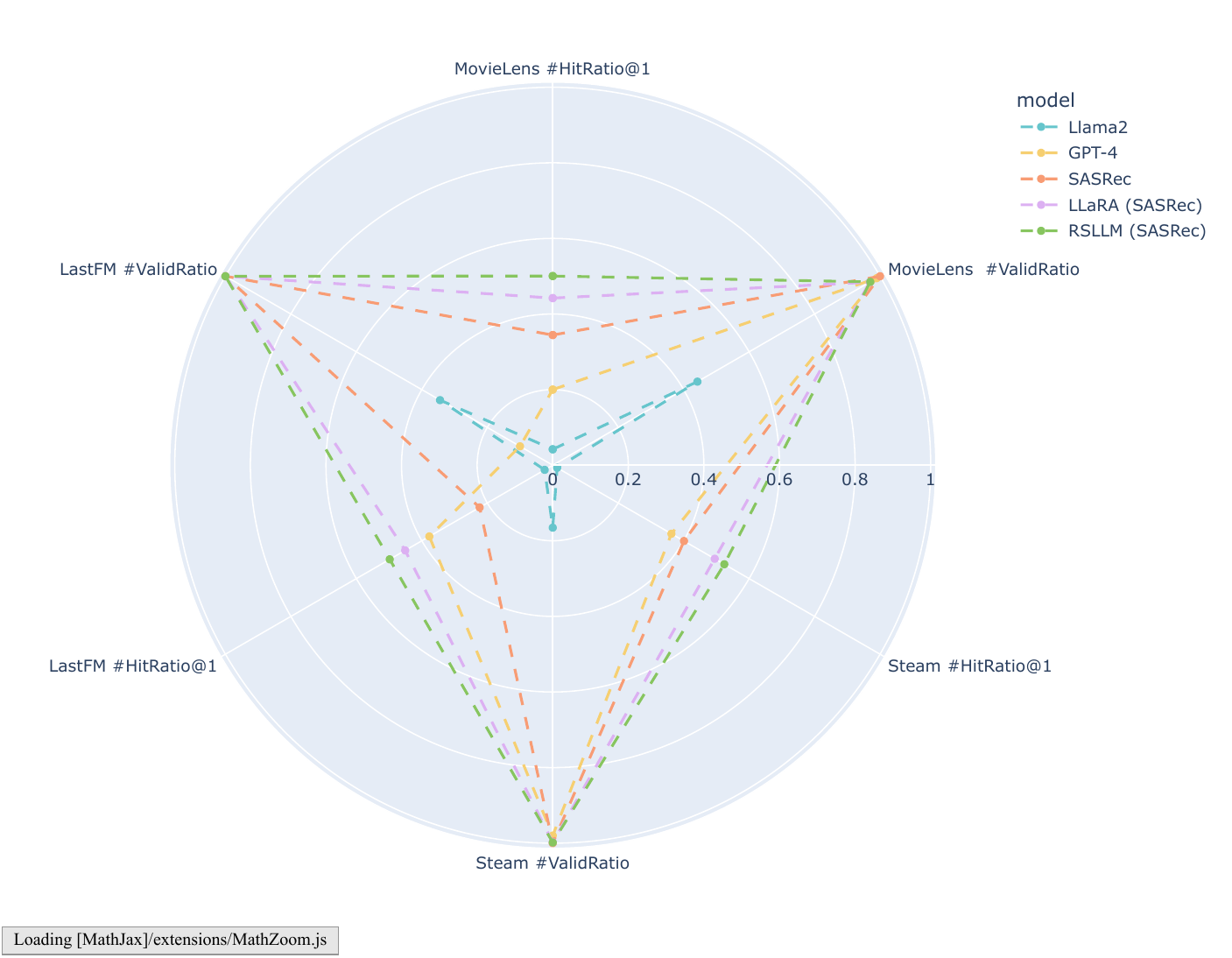}
     \caption{Results of recommendation model efficiency analysis. We compare RSLLM with strong baselines with SASRec backbone.}
     \label{fig:5}
         \vspace{-2em}  

\end{figure}

Figure \ref{fig:4} and Figure \ref{fig:5} presents the results of recommendation model efficiency analysis with GRU4Rec and SASRec in the Figure 3.


\end{document}